\documentstyle[12pt,epsf]{article}             
\begin{document} 
 \title{Growing of the Inhomogeneities with Particle Production}
 \author{ M.de Campos$^{(1)}$ and  N.A.Tomimura$^{(1)}$ . }
\maketitle
   \footnote{
{ \small \it  Instituto de F\'\i sica \vspace{-0.2cm}}\\
{ \small \it Universidade Federal Fluminense  \vspace{-0.2cm}}\\
{\small \it Av. Gal. Milton Tavares de Souza s/n.$\!\!^\circ$, 
                  \vspace{-0.2cm} }}   
\begin{abstract}
Perturbations in Prigogine's cosmological model with particle creation are 
investigated in the framework of Newtonian gravity theory. 
  
The necessary conditions for the density contrast to reach the non-linear regime which is necessary to guarantee
 structure formation are obtained.

The upper limit of the age of the universe is investigated for the perturbations obtained to find them compatible
 with the observed data from COBE(Cosmic Background Explorer satellite).
 
We compare the particle horizon and the Hubble sphere for to guarantee that
 the perturbation is in observable universe.
\end{abstract}
%
%
%
%
%
%
%
%
%
\section{Introduction}

This paper reviews some aspect of our knowledge of the gravitational theory of
 fluctuations of density in homogeneous  and isotropic models of the universe.
 The evolution  of density perturbation is evaluated  for Prigogine's particle
 creation  model for  irreversible  process with  the  framework of  Newtonian
 gravity      theory.       In      this     framework      various      works
 \cite{Gamow} - \cite{Harrison}
 study the time dependence for the  density contrast.  We will find some simple
 conditions on the rate of expansion which permit hydrodynamic perturbation to
 grow rapidally with time.

Here, we just  assume for compatibility with the  cosmological principle, that
any fluctuation  present must  have an amplitude  which decreases  with length
scales or equivalently, mass scale. 

The importance of the reach of  the non linear regime for the density contrast
is relacioned  with the  age of the  structures in the  universe.  Considering
that the structures in the universe to  form ``from top down'', we can use the
period when $\delta _+ \approx 1$ and  the age of the universe to estimate the
age for the structures \cite{Campos}.

%
%
%
%
%
%
%
%
%
\section{Prigogine's Model}
The  universe  has a  considerable  entropy content,  mainly  in  the form  of
blackbody radiation.   Therefore, Einstein equations are  purely adiabatic and
reversible implying  in the  raising of  the question: What  is the  origin of
cosmological entropy?

Prigogine \cite{Prigogine}  proposed a new type of  cosmological history which
includes large scale  entropy production.  These cosmologies are  based on the
reinterpretation  of  the matter-energy  tensor  in  Einstein's equations  and
applied to the homogeneous and isotropic universe, namely

\begin{equation}
      ds^2 = dt^2 -R(t)^2 d\sigma ^2 \, \, ,
\end{equation}
 R=R(t) only.  The Einstein's field equations

\begin{equation}
G_{\mu \nu} = kT_{\mu \nu}
\end{equation}
are  used together with  the adiabatic  transformation in  any system  for the
perfect fluid, namely
\begin{equation}
{T}^{\mu\,\nu}=\left  (\rho+P\right  ){u}^{\mu}{u}^{\nu}+P{g}^{\mu\,\nu} \,  ,
\nonumber
\end{equation}
where
\begin{equation}
P = P_{c}+P_{th} \, .  \nonumber
\end{equation}
$P_{th}$ is the  thermodynamic pressure and $P_{c}$ corresponds  to the matter
creation that is defined by \cite{Prigogine}, namely:
\begin{equation}
P_{c}= - \frac{\rho + P_{th}}{n} (\dot{n} + 3Hn) \, .
\end{equation}

The  process of particle  creation is  to be  consider adiabatic  and pressure
creation  is a  kind  of viscosity  pressure  that is  present  in the  energy
momentum tensor representing the gravitational fluid.  According to the second
law of thermodynamic the particle number variations admitted are such that

\begin{equation}
 dN = d(nV) \geq 0 \, \, .
\end{equation}

This implies  that, in the presence  of matter creation,  the usual Einstein's
equation for (1) is:

\begin{equation}
k\rho = 3H^2 + \frac{\kappa}{R^2}
\end{equation}

\begin{equation}
 \dot{ \rho} = - 3H(\rho + P_{th}),
\end{equation}

to become

\begin{equation}
k\rho = 3H^{2} +\frac{\kappa}{R^{2}} \, \, ,
\end{equation}
and

\begin{equation}
\dot{\rho} = \frac{\dot{n}}{n} (\rho + P_{th}) .
\end{equation}

 In  order  to  exemplify  the  theory  mentioned  above,  they  consider  the
 thermodynamical  pressure as  zero  and  the source  of  the matter  creation
 proportional to $H^2$. Thus

\begin{equation}
 \dot{n} + n\theta = \Psi
\end{equation}
to become
                             \begin{equation}
                              \frac{1}{R^3}  \frac{d}{dt} (nR^3) =  \alpha H^2
                              \, ,
                              \end{equation}
with $\alpha \geq 0$.  This leads to

   \begin{equation}
 R(t)=\{ 1+c(e^{\frac{\alpha k M t}{6}}-1)\} ^{\frac{2}{3}} \, ,
   \end{equation}
                 where
                 \begin{equation}
                           c=\frac{9}{kM\alpha}
                           (\frac{kMn_{0}}{3})^{\frac{1}{2}} \,
                  \end{equation}
  $M$ is the mass of particle produced and $n_{0}$ is the initial number
  density of particles.
                  
 The evolution of the scale factor  is similar to the models with cosmological
 term \cite{Frieman} and have the pattern demonstrate in fig.1.
\begin{figure}[!h]
\epsfysize=12.cm \centerline {\epsfbox{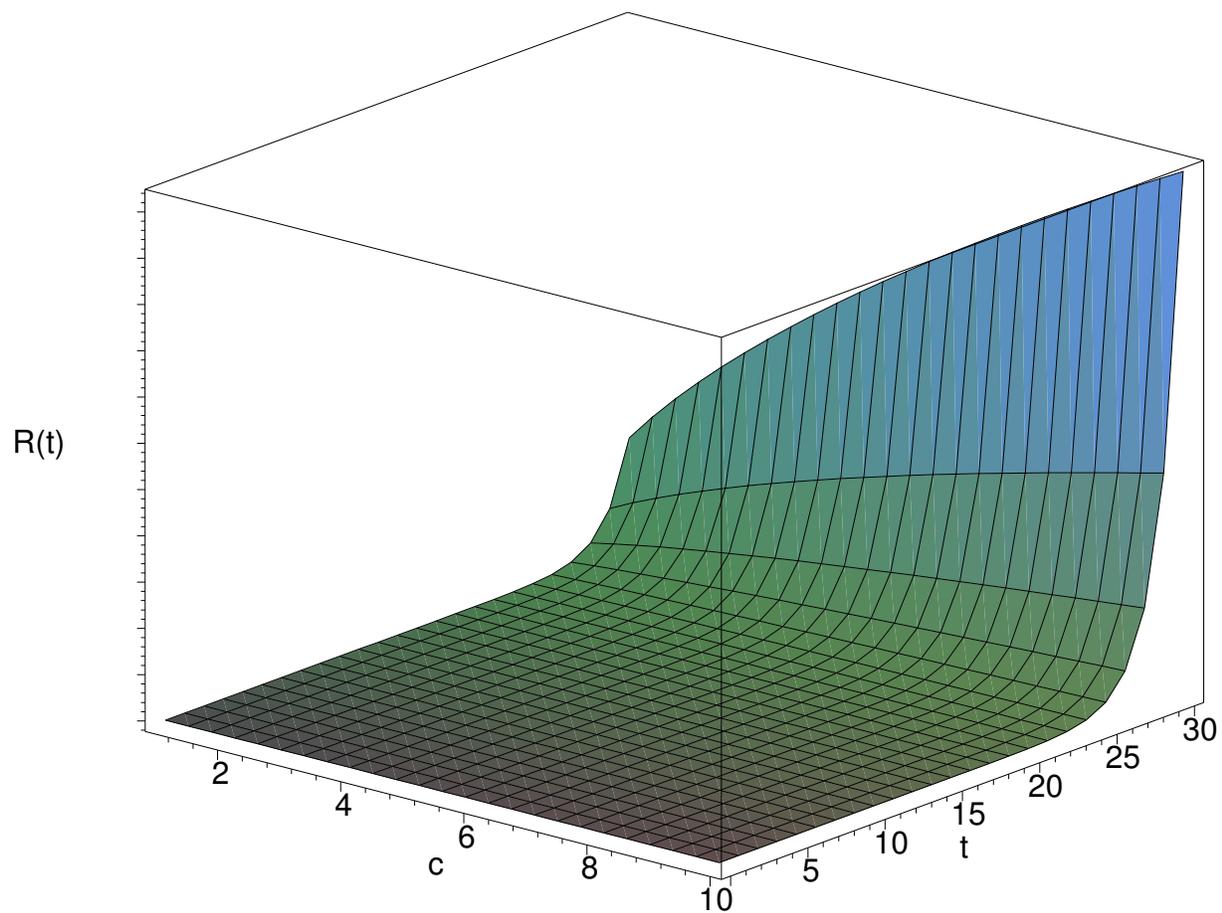}}
\caption{Evolution for the scale factor $ \bf{R(t)}\times \bf{t} \times c$ \, .}
\label{fig:figura 1}
\end{figure}
\newpage Note that  the recent measurement of distant  supernova when they are
used   as   standard  candles   \cite{Riess}   \cite{Pelmuter}  indicates   an
accelerating   universe,  consequently   the  desaceleration   parameter  will
establish  restrictions  on  the  creation parameters  of  Prigogine's  model.
Namely
\begin{equation}
q=-\frac{\ddot{R}R}{\dot{R}^2}      \Longrightarrow      q=\frac{1}{2c}\{-2c+3
(c-1)e^{(-\frac{\alpha k M}{6} t)}\} \, .
\end{equation}
For an  accelerating universe  $q$ must be  negative.  This  condition imposes
 restrictions for the time coordinate which is:
\begin{equation}
t> - \frac{6}{\alpha k M}\ln{\frac{2c}{3(c-1)}}\, .
\end{equation}
If we  wish to have  a positive time  , consequently $c>3$.  Therefore,  if we
don't  truncate, arbitrarily  second  Abramo and  Lima  \cite{Lima}, the  time
coordinate  at $t=0$,  then the  only restriction  is given  by  the condition
(16).
 %
 %
 %
 %
 %
 %
 %
 %
 %
\section{Differential equation for contrast evolution}  
The  fundamental hydrodynamical  equations  that describe  the  motion of  the
cosmic fluid here considered are

\begin{equation}
{(\partial \vec{u})  _r \over (\partial  t)} +(\vec{u}\cdot \vec{\nabla  _r })
\vec{u}= -\vec{\nabla _r} \Phi-{\vec{\nabla _r } P \over \varrho}
\end{equation}

\begin{equation}
({\partial \rho \over \partial t })_r+ \vec{\nabla } _r(\rho u)=\Psi
\end{equation}

\begin{equation}
\nabla _r^2\Phi=4\pi G\varrho \, .
\end{equation}

The sub-script $r$ refers to the proper distance referential.

It should be observed that the particle creation has modified Euler's equation
\cite{Waga}, namely

\begin{equation} 
\vec{\nabla } P _{c} = \Psi (\vec{u}_{p}-\vec{u} ) \, \, .
\end{equation}

It will be assumed that the velocity of the created particles $ \vec u_{p}$ is
  the   same  as  the   already  existing   ones.   Equations   (17)-(19)  are
  respectively,  the   momentum  conservation  law   (Euler's  equation),  the
  continuity equation and Poisson's modified equation.
 
 Here $\vec{u}$ is  the velocity of a fluid element, $\rho$  is the fluid mass
 density, $\Phi$  is the  gravitational potential ,P  is the total  pressure .
 The  fluid  we  consider  here  is  a cold  one,  such  that  $P_{th}=0$  and
 $P_{c}=P_{c}(t)$ only.
 
 We  now introduce  the comoving  coordinate $\vec{x}$  related to  the proper
 coordinate $\vec{r}$ by \cite{Peebles}

\begin{equation} 
 \vec{r}=R(t)\vec{x} \, \, ,
\end{equation}
 where  $R(t)$  is  the expansion  factor.   Equation (21)  is a  change  of
variables  from  proper   locally  Minkowskian  coordinates  $(\vec{r},t)$  to
expanding  coordinates $(\vec{x},t)$  comoving  in the  background model.   In
these latter  coordinates the  proper velocity of  a particle relative  to the
origin is

\begin{equation} 
\vec{u}    =   a    \vec{\dot{x}}+\vec{x}   \dot{a}    =    \dot{a}\vec{x}   +
\vec{v}(\vec{x},t) .
\end{equation}

If a small  perturbation is placed on the  background, first order corrections
appear  in  the   velocity,  density  and  potential.   We   will  call  these
$\vec{v}(\vec{x},t)$, $\delta  (\vec{x},t)$ and $\phi  (\vec{x},t)$.  We shall
assume that these quantities are small such that $\delta << 1$ and $\vec{v} <<
\vec{u}$.  Peebles \cite{Peebles}  shows that the potential $\phi$  in the new
coordinates become
 
 \begin{equation} 
 \Phi = \phi -\frac{1}{2} R \ddot{R} x^2 \, .
\end{equation}
We shall not detain to derive equations (17) - (19) in the new coordinates. We
list them only and refer the reader to \cite{Peebles}
\begin{equation}
\nabla ^2 \phi = 4\pi G R^2 (\rho +3P)
\end{equation}

\begin{equation}
\frac{\partial  \rho}{\partial  t}  -\frac{\dot{R}}{R} \vec{x}  .  \vec{\nabla
}{\vec{u}} +\frac{(\vec{u}.\nabla )\vec{u}}{R} = \Psi
\end{equation}

\begin{equation}
R\frac{\partial \vec{u}}{\partial t} -\dot{R}\vec{x} . \vec{\nabla }{\vec{u}}+
\frac{\vec{u}.\vec{\nabla }{\vec{u}}}{R} = \frac{\vec{\nabla }\phi}{R} \, .
\end{equation}

Expanding $\rho$  , $\vec{u}$  and $\phi$ pertubatively  and keeping  only the
first order terms in the  equations above gives the linearized equations.  The
perturbations are defined as follows

\begin{equation}
\rho = \rho _{0} (t) (1+\delta (x,t))
\end{equation}

\begin{equation}
\vec{u} = R\vec{\dot{x}} + \vec{v}
\end{equation}

\begin{equation}
\Phi = \phi  - \frac{1}{2}R\ddot{R}x^2 = \phi +  \frac{2}{3} \pi G\rho_{0} R^2
x^2 +2P_{c}R^2 x^2 \pi G \, ,
\end{equation}

where
\begin{equation}
\mid \vec{v} \mid \, , \mid \phi \mid \, , \mid \delta \mid \, << 1 \, .
\end{equation}
The field equation of zero order has been used, namely
\begin{equation}
{3{\ddot R} \over R}=-4\pi G \rho_0 -12\pi G P_{c} \, .
\end{equation}

The linearized perturbative equations become

\begin{equation}
\nabla^2 \phi=4 \pi G {R^2} \rho _{0} \delta
\end{equation}

\begin{equation}
\frac{\partial   \vec  v}{\partial  t}   +  \frac{\dot   R}{R}  \vec{v}   =  -
\frac{\vec{\nabla } \phi}{R}
\end{equation}

\begin{equation}
\vec{\nabla  }  .  \vec   v  =  -R  [\frac{\partial  \delta}{\partial  t}+\Psi
\frac{\delta}{\rho_{0}}].
\end{equation}

From these  equations we can obtain  a second order  differential equation for
the linearized  density contrast, $\delta$,  by taking divergence  of equation
(33) together with (32) and (34) we get
\begin{equation}
\ddot \delta +\{2\frac{\dot  R}{R}+\frac{\Psi}{\rho_0}\}\dot \delta - \{4\pi G
\rho_0      -2\frac{\dot     R}{R}\frac{\Psi}{\rho_0}-\frac{\partial}{\partial
t}(\Psi/\rho_0)\}\delta = 0\, .
\end{equation}

In order to integrate equation (35),  it will be convenient to change variable
from $t$ to $U$ such that
\begin{equation}
R(t) = U^{\frac{2}{3}} \, .
\end{equation}

After some algebra equation (35) is transformed into
\begin{equation}
U^{2}(U-a)\delta^{''}   +\{\frac{4}{3}    (U-a)U+(k_3   +1)U^2   \}\delta^{'}-
\{\frac{2}{3} (U-a) - \frac{4}{3} k_3 U\} \delta = 0 \, ,
\end{equation}
where  $k_3=\frac{2}{M}$ and  $a =  1 -  c $.   The derivative  is  taken with
  respect to  the variable $U$.  By  integrating the above  equation we obtain
  the following hypergeometric functions
\begin{eqnarray}
\delta    _1=U^{2/3}\{Y(\alpha_{1},\beta_{1},\gamma_{1},U/a)\}    \\    \delta
_2=U^{-1}\{Y(\alpha_{2},\beta_{2},\gamma_{2},U/a) \} \, ,
\end{eqnarray}
where
\begin{eqnarray}
\alpha_{1}=\frac{8+3k_{3}+\sqrt{9k_{3}^{2}-24k_{3}+40}}{6}\\
\beta_{1}=\frac{8+3k_{3}-\sqrt{9k_{3}^{2}-24k_{3}+40}}{6}\\    \gamma_{1}=8/3\\
\gamma_{2}=-2/3\\
\alpha_{2}=\frac{+3k_{3}-2+\sqrt{9k_{3}^{2}-24k_{3}+40}}{6}\\
\beta_{2}=\frac{+3k_{3}-2-\sqrt{9k_{3}^{2}-24k_{3}+40}}{6} \, .
\end{eqnarray}
It  follows that $\delta  (U)$ represents  a family  of solutions  of equation
(37), labeled by the parameters $\alpha _1$, $\beta _1$,$\alpha _2$ and $\beta
_2$.   It might  be noted  that the  behavior  of $  \delta $  in this  linear
approximation depends  only on the initial  values of $ \delta  $ and $\frac{d
\delta}{dU}$.
%
%
%
%
%
%
%
%
%
\section{Growing and decaying modes}
In order to have  an insight about the nature of the  solution we consider the
case where $k_3 =1$ and  $c>1$, such that the hypergeometric solutions $\delta
_{+}$ and $\delta _{-}$ become

\begin{eqnarray}
\delta_{-}=\frac{c_1}{U(U-a)}\\ \delta_{+}=\frac{c_{2}U^{2/3}}{U-a} \, \, ,
\end{eqnarray}
where  $c_1 $  and $c_2  $  are integration  constants, $\delta  _{+}$ is  the
growing mode and $\delta _{-}$ is the decaying mode.

Although the decreasing mode can  be important in some circumstances, we shall
hereafter mainly deal only with the increasing mode. It is responsable for the
formation   of  the   cosmic  structure   in  the   gravitational  instability
picture. Beyond this the universe would  not have been homogeneous in the past
if we consider the coefficient of the decaying mode different of zero.

Substituting (13)  in (17), we  can  express  the mass  density
contrast as a function of the parameter $\tau $, namely
\begin{equation}
\delta _{+} = w[e^{- \tau }-(1-\frac{1}{c}) e^{-3 \tau}]^{\frac{2}{3}} \, ,
\end{equation}
with $w=c_2 c^{\frac{1}{3}}$ and
\begin{equation}
\tau = \frac{\alpha k M t}{12}\, .
\end{equation}

The profile for the $\delta _ + $ is given below

\begin{figure}[!ht]
{\epsfxsize=10.cm {{\epsfbox{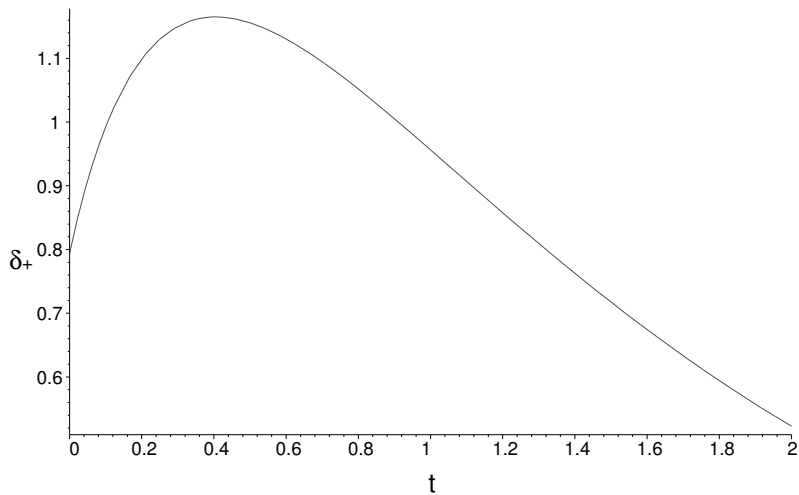}}}}
\caption{Evolution of  the growing mode  for the density contrast  when $c=4$.
  For mathematical convenience we take $w=2$.}
\end{figure}
\newpage and for $c = 10$ by
\begin{figure}[!ht]
{\epsfxsize=10.cm {{\epsfbox{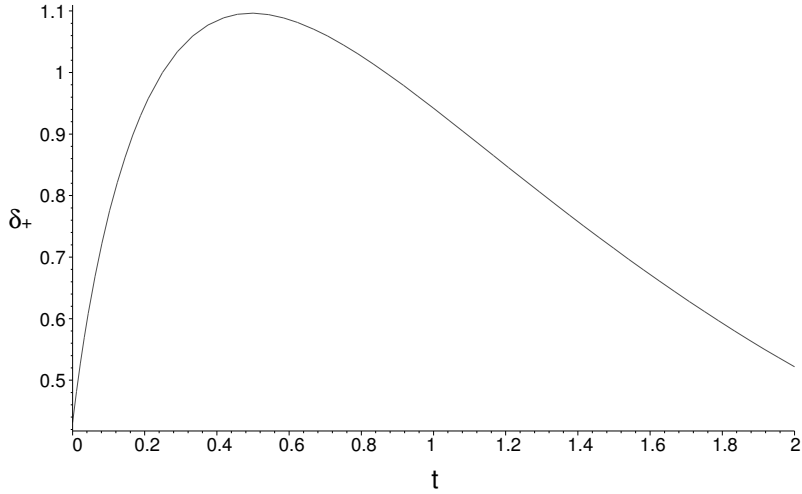}}}}
\caption{Evolution of the growing mode for the density contrast when $c=10$. 
  For mathematical convenience we take $w=2$. }
\end{figure}
%
%
%
%
%
%
\section{The reach of the non linear regime for the density contrast}
In order  to evaluate the  time for the  perturbation to reach  the non-linear
regime for the density contrast in our model, we use the Sachs-Wolf
relation together with equation (48). Thus
\begin{equation}
\frac{\delta T}{T} \propto \frac{\delta \rho}{\rho}
\end{equation}
is applicable for the considered models.

It is known that \cite{Kolb}
\begin{equation}
(\frac{\delta \rho}{\rho})_{dec} \leq O(10^{-2} - 10^{-3}).
\end{equation}
If we insert these  values in our model, that is, in  expression (48) one gets
for $\tau$ the following
\begin{equation}
\tau _{dec} \sim O(10^{-4} - 10^{-5}) \, ,
\end{equation}
  where $\tau _{dec}$  corresponds to the parameter at the decoupling time.

It seems reasonable to suppose  that the present strong condensations grew out
of  small  disturbances  so  that  a necessary  (if  perhaps  not  sufficient)
condition  for  their  formation   is  that  the  perturbation, $\delta  _{+}$
, calculated in linear stability should have  become of order unity at some time
before the present \cite{Weinberg}.

 Consequently $\delta _{+} \sim 1 $ implies that
\begin{equation}
\tau _{RL} \sim 10^{-1}
\end{equation} 
in our model.


We construct the quocient
\begin{equation}
Q = \frac{\tau _{RL}}{\tau _{dec}} = \frac{t_{RL}}{t_{dec}} \, 
\end{equation}
for to determinate the time for  the perturbations reach the non linear regime
($ t _{RL}$).  Using (53) and (54) we obtain
 \begin{equation}
Q = \frac{\tau _{RL}}{\tau _{dec}} \sim O(10^{3} - 10^{4}) \, .
\end{equation}
If we  insert these values into (54)  and take in account  that the decoupling
occurred at $t \sim 10^{6} ys $ one finds
\begin{equation}
t_{RL} \sim O(10^9 - 10^{10}) \, .
\end{equation}

%
%
%
%
%
%
%
%
\section{The horizon problem}
Another crucial problem is to verify the horizon problem for Prigogine's model.  To this end, one should point out 
the distinction between the cosmological particle horizon and the Hubble sphere, or speed of light sphere, $R_c $, which is 
simply defined to be the distance from $\bf{O}$ of an object moving with the cosmological expansion and the velocity 
of light with respect to $\bf{O}$ .  This can be seen very easily to be
\begin{equation}
R_{c} = \frac{R}{\dot{R}} = \frac{3}{4} \{ \frac{1+c(e^{2\tau }-1)}{ce^{2\tau
    }} \} \, \, .
\end{equation}

The  cosmological particle horizon at time $ t $ is given by
\begin{equation}
R_{H} = R(t) \int _{0} ^{t} {R(t^{'})^{-1}dt^{'}} \, \, .
\end{equation}

For our model, this is given by
\begin{eqnarray}
R_{H}                  &=&                  \frac{1}{4}(b-U)^{\frac{2}{3}}[\ln
{\frac{(U^{\frac{1}{3}}+b^{\frac{1}{3}})   (1-b^{\frac{1}{3}}+b^{\frac{2}{3}})}
{(U^{\frac{2}{3}}-(bU)^{\frac{1}{3}}+b^{\frac{2}{3}})
(1+b^{\frac{1}{3}})^{2}}}                      -2                     \sqrt{3}
\arctan{\frac{\sqrt{3}}{3}(\frac{-2U^{\frac{1}{3}}+b^{\frac{1}{3}}}{b^{\frac{1}{3}}})}+\nonumber
\\                                                                 &+&2\sqrt{3}
\arctan{\frac{\sqrt{3}}{3}(\frac{b^{\frac{1}{3}}-2}{b^{\frac{1}{3}}})}\, .
\end{eqnarray}
 $U$  is given  by  (36), $b  = c-1$  and  c is  given  by (14).   Due to  the
analytical difficulties of comparing $R_{c}$  and $R_{H}$ we plot their graphs
(fig.4).  The  horizon problem is solved when  $c$ is large (that  is when the
initial number of particles per volume is large).  \\
\begin{figure}[!ht]
\begin{tabular}{ccc}
{\epsfxsize=4.cm {{\epsfbox{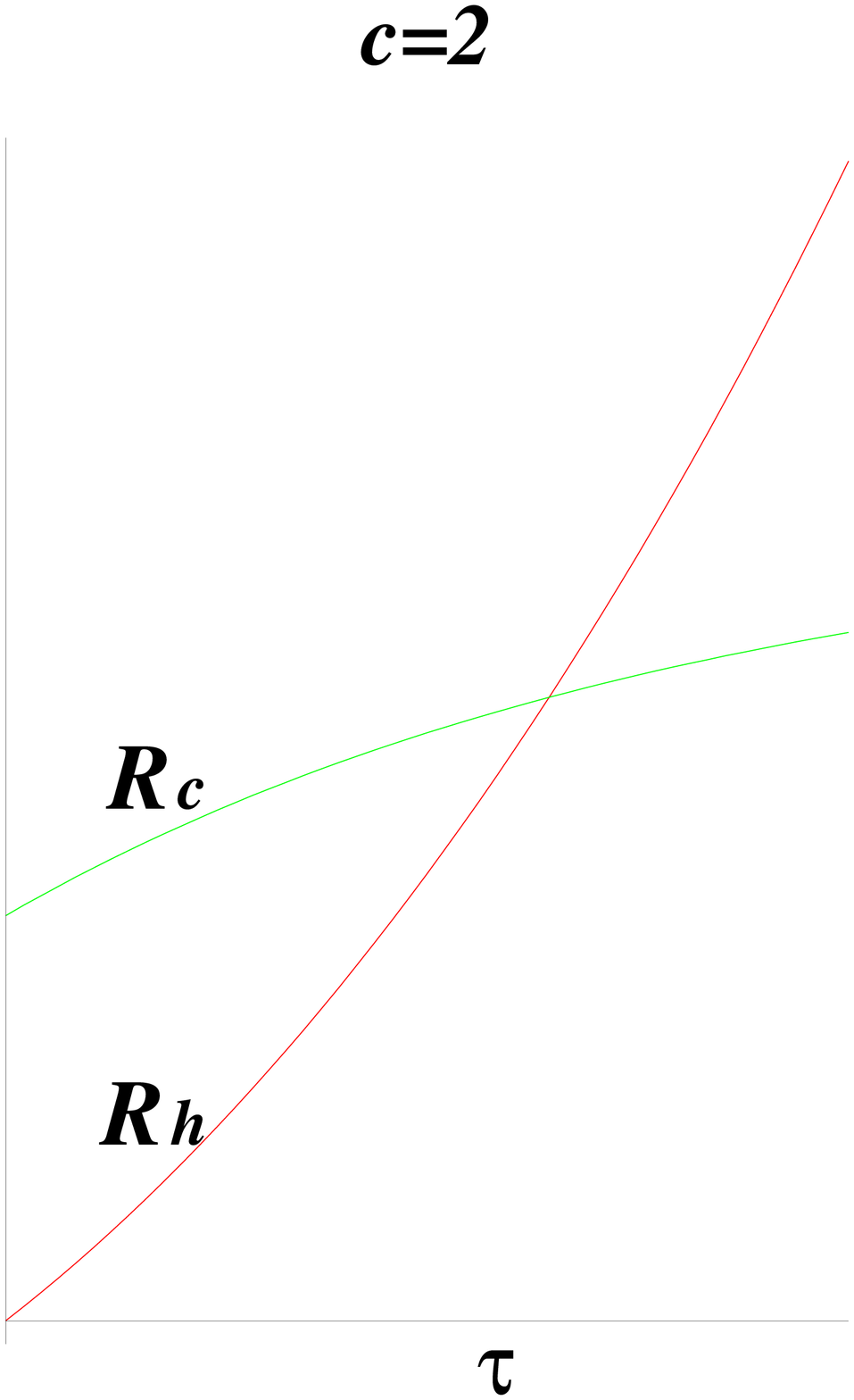}}}} &
{\epsfxsize=4.cm {{\epsfbox{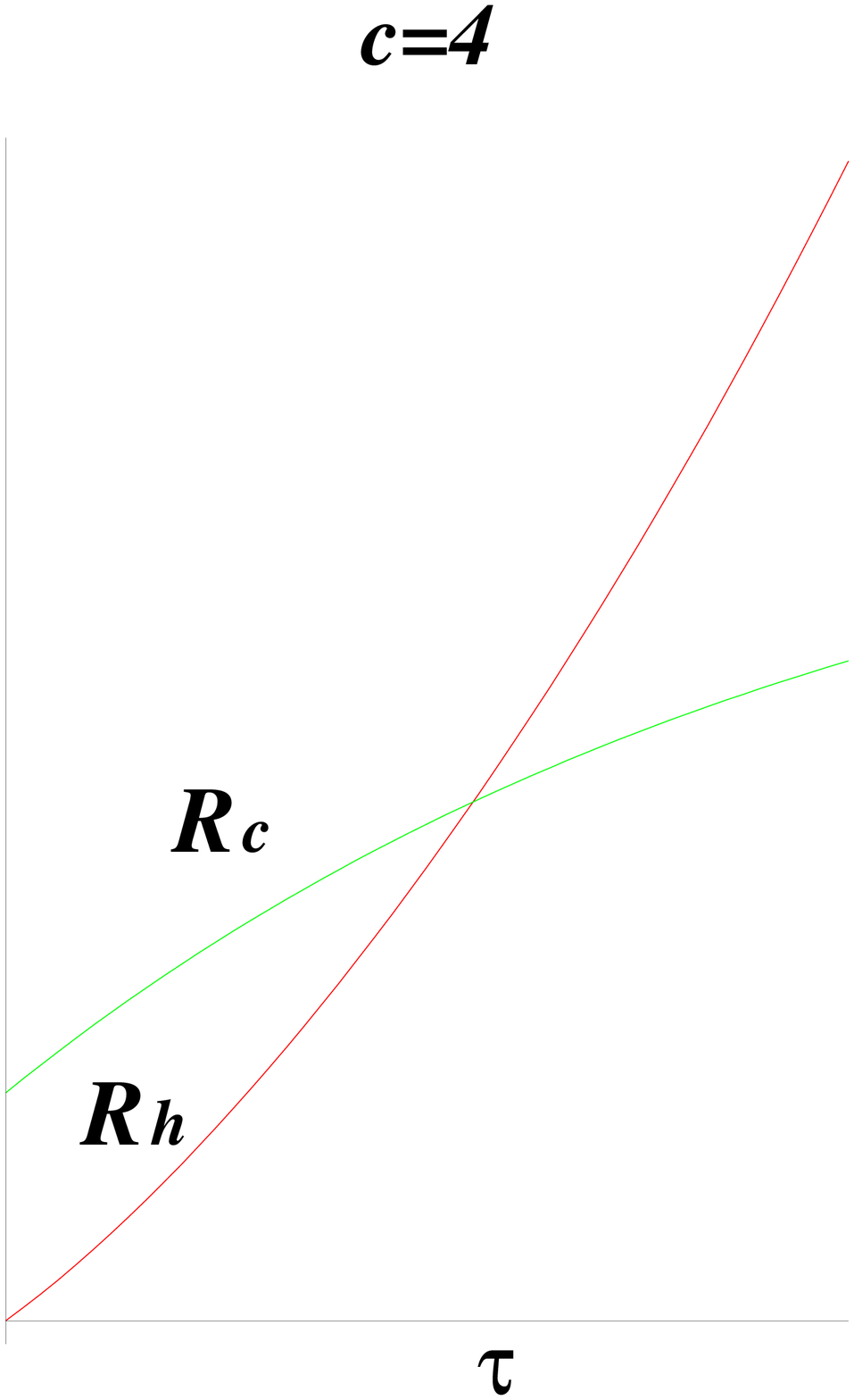}}}} &
{\epsfxsize=4.cm {{\epsfbox{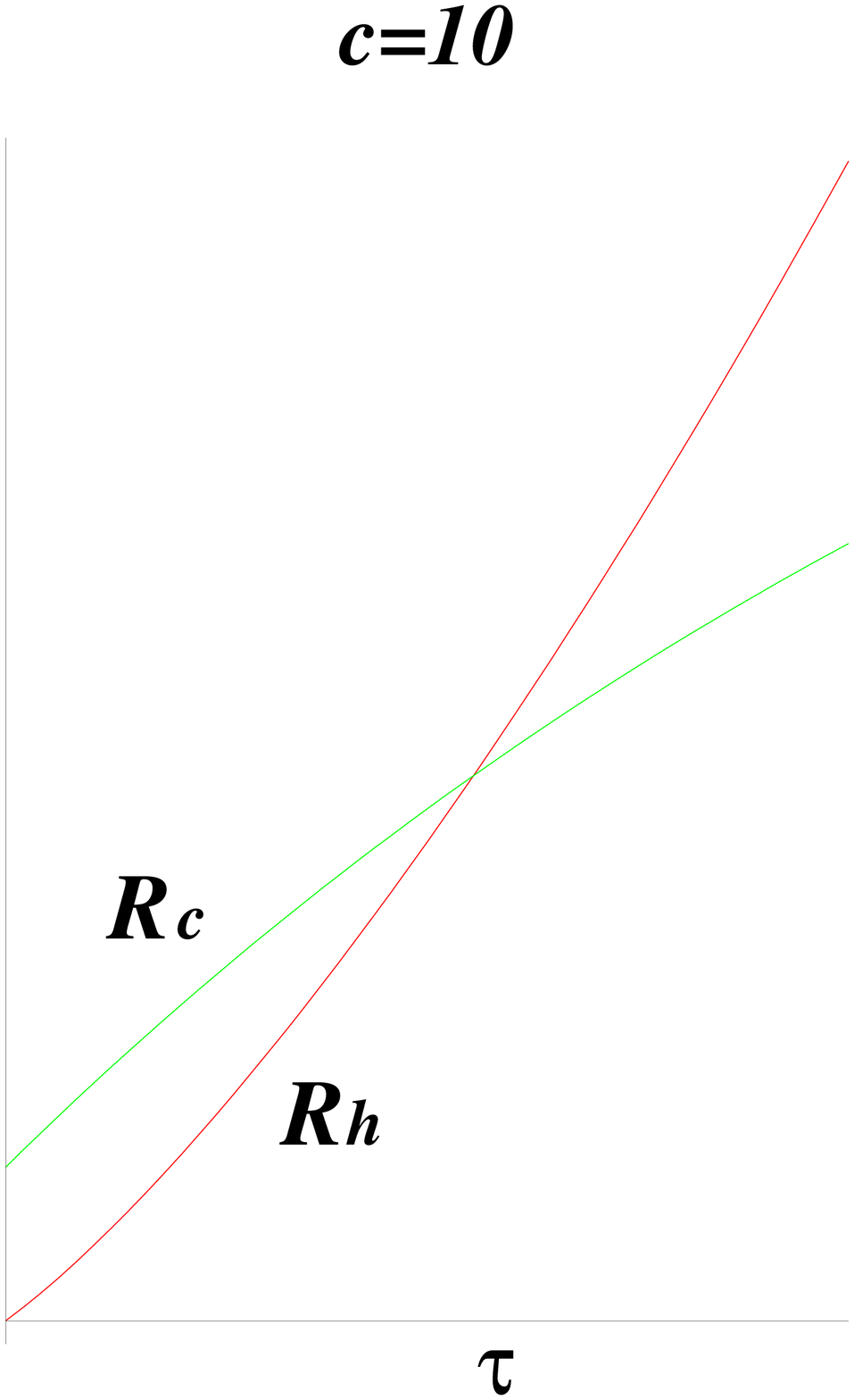}}}} 
\end{tabular} 
\end{figure}
\newpage
\begin{figure}[!ht]
\begin{tabular}{ccc}
{\epsfxsize=4.cm {{\epsfbox{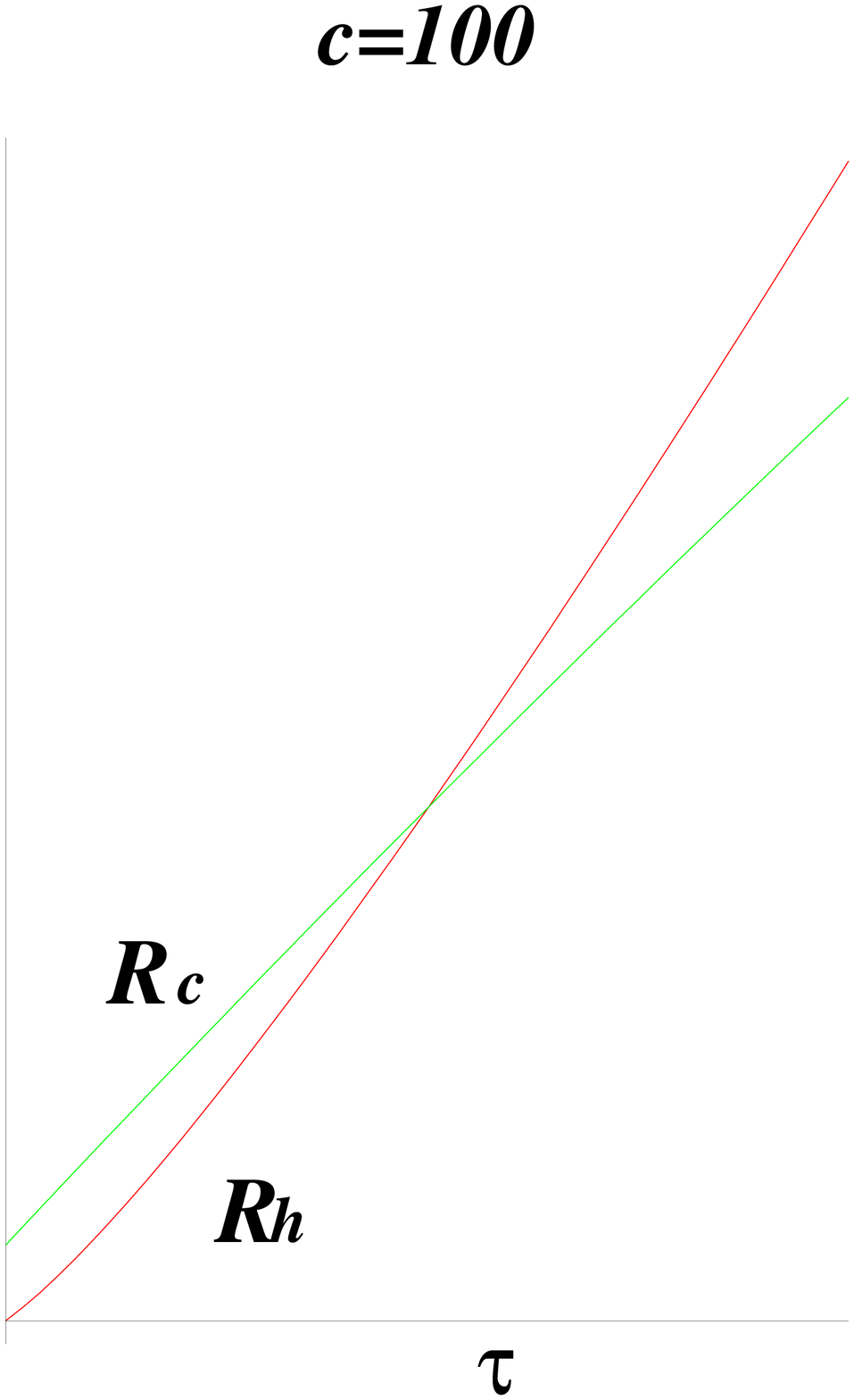}}}} &
{\epsfxsize=4.cm {{\epsfbox{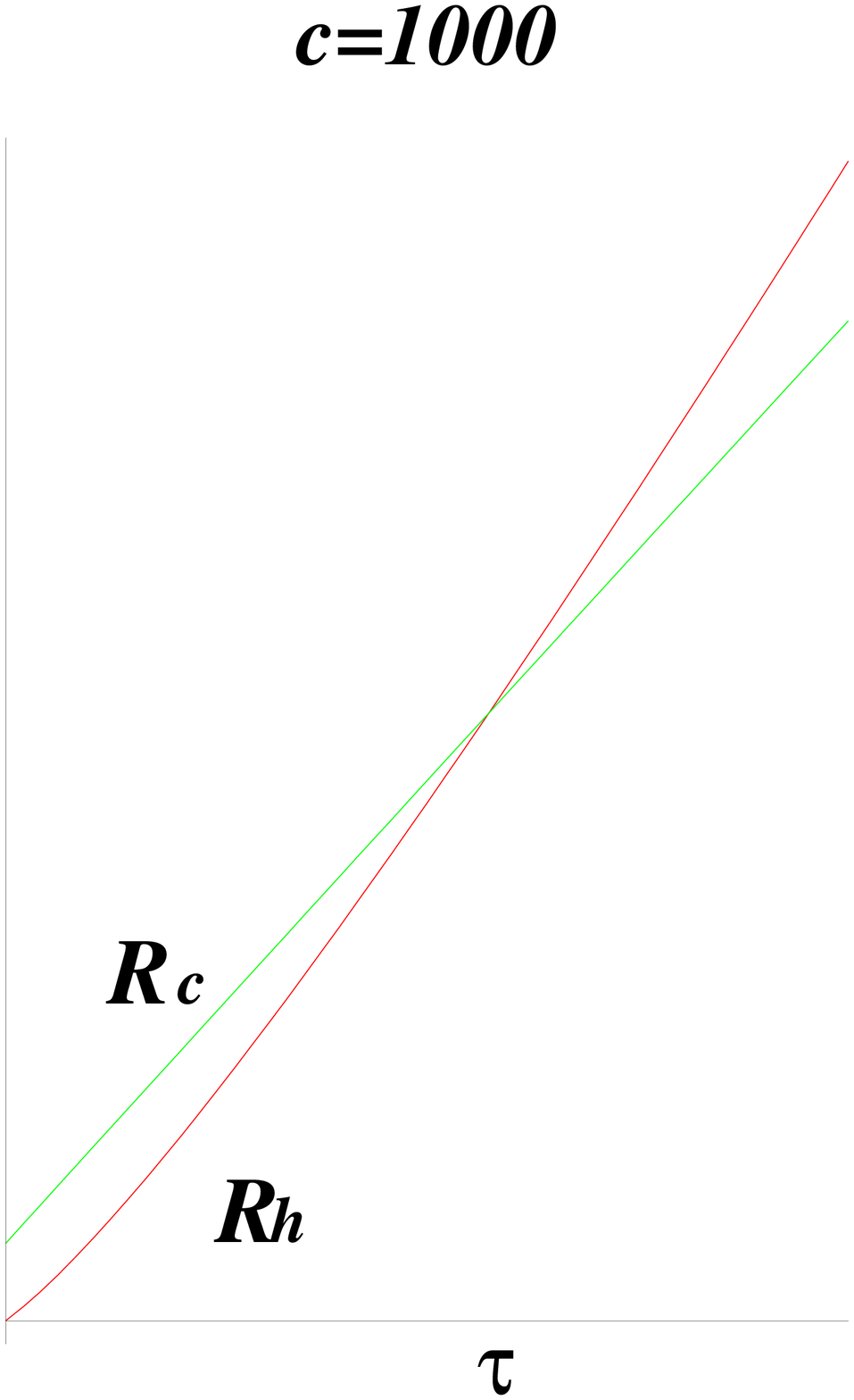}}}} &
{\epsfxsize=4.cm {{\epsfbox{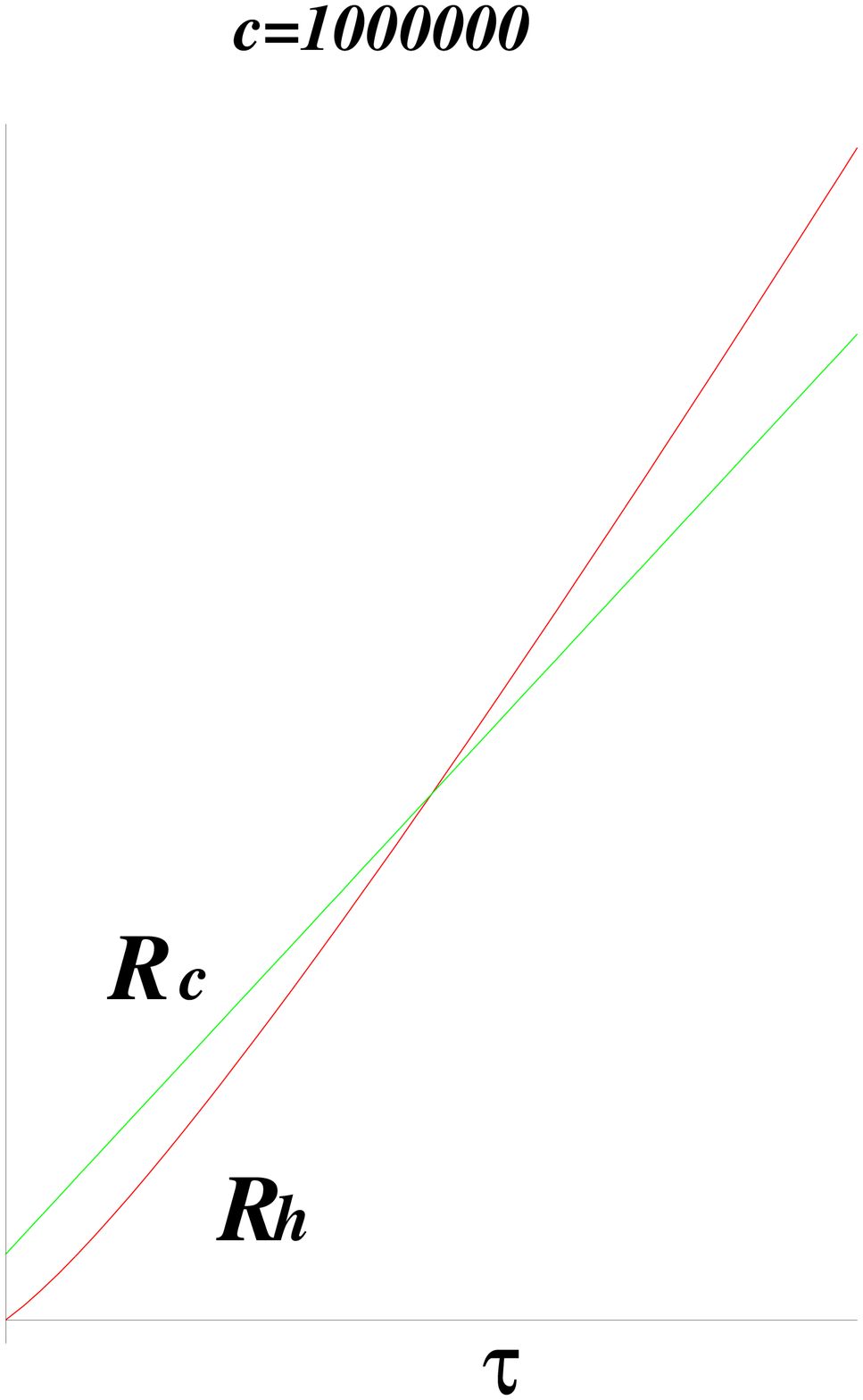}}}} 
\end{tabular}
\caption{Cosmological particle horizon(red) and Hubble sphere(green) vs. $\tau$ for various values of $c$. 
The particle horizon intercept the Hubble sphere for: $c=2 \rightarrow \tau \sim 0.38$; $c=4 \rightarrow \tau \sim 0.27$; 
$c=10 \rightarrow \tau \sim 0.16$; $c=100 \rightarrow \tau \sim 3*10^{-2}$; $c=1000 \rightarrow \tau \sim 3*10^{-3}$; 
$c=1000000 \rightarrow \tau \sim 3*10^{-6}$ \, .}
\end{figure}
%
%
%
%
%
%
%
%
%
\section{The Age of the universe}

We now turn our attention to evaluate the characteristic time scale for the
age of the universe with the ultimate aim of determining $t_{p}$, the time
that elapsed from the beginning of the universe until now.

The recent results from the Cosmic Background Explorer (COBE) strongly
supports the gravitational instability theory of structure formation and can
be interpreted as providing evidence for the existence of small
inhomogeneities generated in the early era.  For the Priogogine's model the
perturbation is given by
\begin{equation}
\delta _{+} = \frac{R(t)}{e^{\beta t}}  \, ,
\end{equation} 
where
\begin{equation}
\beta =\frac{\alpha k M }{6} \, .
\end{equation} 

With the aid of equation (60) it can be written as
\begin{equation}
\frac{\delta _{+} (t_{p},\beta)}{\delta _{+} (t_{dec},\beta)} = (1+Z_{dec}) e^{-\beta (t_{p}-t_{dec})} \, ,
\end{equation}
where $Z_{dec}$ is the decoupling redshift.

By assuming that during the matter dominated era particle production are restricted to non relativistic ones and its amplitude is fixed at decoupling time
(COBE roughly fix this amplitude), we have
\begin{equation}
\frac{\delta _{+} (t_{p},\beta)}{\delta _{+} (t_{p},\beta = 0)} = e^{-\beta (t_{p} -t_{dec})} = \frac{\sigma_{8} (\beta)}
{\sigma_{8} (\beta = 0)}
\end{equation}
and 
\begin{equation}
\delta_{+} (t_{dec} , \beta) = \delta_{+} (t_{dec}, \beta = 0) \, ,
\end{equation}
where $\sigma_{8} (\rho)$ is the root-mean-square mass fluctuation in sphere of radius $8h^{-1} MPc$.  The recent COBE measurements
indicate that $\sigma_{8} (\beta = 0) \sim 1$ for the standard cold dark matter \cite{Liddle}.

By requiring that structures form not to late, it is safe to assume
$\sigma_{8} (\beta) \geq \frac{1}{3}$ \cite{Lyth}.  Using this condition into 
equation (63) implies that
\begin{equation}
e^{-\beta (t_{p}-t_{dec})} \geq \frac{1}{3} \, .
\end{equation}

If we assume that $t_{p}>>t_{dec}$, a superior limit for the age of the universe may be presented as
\begin{equation}
t_{p} \leq \frac{6.6}{\alpha k M} \, ,
\end{equation}
which becomes in term of the parameter $\tau$, using (49), as
\begin{equation}
\tau _{p} \leq 0.55 \, .       
\end{equation}

In order to give a better idea on the superior limit of the age of the universe, we follow the same procedure used in the last section.
We construct the quotient
\begin{equation}
\tilde{Q} = \frac{\tau _{p}}{\tau _{dec}} \, .
\end{equation}
Fixing $\tau _{p} \sim 0.55 $ and $\tau _{dec} \sim 10^{-4} - 10^{-5}$, one gets 
\begin{equation}
 \tilde{Q} \sim O(10^4 - 10^5) \, .
\end{equation}
Once again if we take for the decoupling time $t_{p} \sim 10^6$ ys one has then
\begin{equation}
t_{p} \sim (10^{10} - 10^{11})  \, \, ys \, .
\end{equation} 
This result indicates that for the Prigogine's model, the age of the universe is superior to the age of the structure 
formation and have an interval which agrees with the actual estimation.
%
%
%
%
%
%
%
%
\section{Conclusions}

In summary, in this paper we analyze the growth of density perturbation in Newtonian cosmological model for Prigogine's
type model to find a class of hypergeometric solutions.

 A special case is studied to obtain an increasing  and a decreasing mode analogous to Friedman model.
 
 An estimate of the time interval for the perturbation to reach the non-linear regime is obtained.
 
 An upper limit of the age of the universe is obtained by making use of Cobes' data.
 
 The horizon problem was carefully explained to show that for big values of
 the parameter $c$ the space-time is connected casually.

\section*{Acknowledgments}
This work was supported by CNPq (Brazilian Agencie).

\end{document}